\documentclass[twocolumn,showpacs,showkeys,preprintnumbers,superscriptaddress, pra, 10pt, aps]{revtex4-2}
\usepackage[utf8]{inputenc}
\usepackage[english]{babel}
\usepackage[caption=false]{subfig}
\usepackage{graphicx}
\usepackage[T1]{fontenc}
\usepackage{listings}
 \usepackage{vwcol}
 \usepackage{mathtools}
 \usepackage[]{booktabs}
\usepackage{amsmath}
\usepackage{amssymb}
\usepackage{diagbox}
\usepackage{float}
\usepackage[colorlinks=true,linkcolor=blue,urlcolor=blue,citecolor=blue]{hyperref}
\usepackage[capitalize]{cleveref}  % Intelligent references...

\usepackage{fixme}
\usepackage[colorinlistoftodos]{todonotes}

\fxsetup{status=draft, inline=true, margin=false} % <====== add this line

\begin{document}

\renewcommand{\lstlistingname}{Código}
\renewcommand{\lstlistlistingname}{Índice de fragmentos de código fuente}
\newcommand{\reflabel}{\mathrm{c}}
\newcommand{\omegac}{\omega_\reflabel}
\newcommand{\charlen}{a_\reflabel}
\newcommand{\gpeham}{\hat{H}_\mathrm{GP}}
\newcommand{\extpotential}{V_\mathrm{ext}}
\newcommand{\chardensity}{n_\reflabel}

\newcommand{\NormPsi}{\Psi'}
\newcommand{\ImagTimePsi}{\tilde{\Psi}}
\newcommand{\planeWaveAmplitude}{\tilde{\psi}_k}

\title{%Ground-State Properties for a Bose Gas Within a One-Dimensional Imperfect Crystal \\
Vacancy Effect on the Ground-State Energy of a Bose Gas 
\\Trapped by 1D Imperfect Artificial Crystal}
\author{E.I. Guerrero-Cruz}
\affiliation{Posgrado en Ciencia e Ingeniería de Materiales, UNAM}
%\author{O.A. Rodríguez-López}
\author{M.A. Sol\'is }
\affiliation{Instituto de F\'isica, Universidad Nacional Aut\'onoma de M\'exico,
  Apdo. postal 20-364, 01000 Ciudad de México, Mexico}
%	juan.valencia@uacm.edu.mx,  masolis@fisica.unam.mx}

\date{Modified: \today / Compiled: \today}

\begin{abstract}
   For a weakly interacting Bose gas trapped by an imperfect one-dimensional artificial crystal, we study the effect of its punctual defects, i.e. vacancies, on the ground state properties of the system. In the framework of the mean field approximation, we numerically solve the corresponding Gross-Pitaevskii equation using the ``Gradient Flow with Discrete Normalization'' method, also known as the imaginary time method. The crystal is artificially produced by applying an external Dirac comb potential to the Bose gas where vacancies are created by randomly removing a predetermined number of deltas. We observe that as the number of randomly removed deltas increases, the ground state energy decreases exponentially from its value for the perfect crystal case until it reaches its value when the Bose gas is free.
    The ground state energy is reported for different magnitudes of the interaction between bosons and several system sizes which we extrapolate to infinity for the crystal with only one vacancy. Also, we observe the presence of an energy gap between the ground state energies of the perfect system and that with a vacancy, which is more noticeable for values of the particle interaction magnitude $ g \leq 0.1$, when the delta strength $P_0 = 10$. In addition, we report the boson distributions within the crystal,
    %inside a box with periodic boundary conditions,
     i.e. the probability density functions which show
    localization features around vacancies which disappear as $g$  increases. From the ground state energy, the chemical potential is obtained immediately.

\end{abstract}

\maketitle

\section{Introduction}

We know that Bose-Einstein condensation (BEC) of a free ideal Bose  gas (IBG) occurs at a non-zero temperature when the dimension $d$ of the gas is greater than the power of the
momentum magnitude $k$ in its energy-momentum relation, e.g., for the 3D IBG $\varepsilon = \hbar^2 k^2 /2m$ 
and  $d > 2$ \cite{Solis}. However, when
the IBG is trapped by an external potential its energy-momentum relation could be different from quadratic and BEC could occur at non-zero temperature for $d \leq 2$
 \cite{Faruk, Bagnato}. If the external potential is a {\it perfect} periodic potential like a Dirac comb, the energy dispersion relation presents a structure of allowed and prohibited energy bands which does not modify the criterion for the appearance of BEC, at a temperature different from zero, with respect to that of the free IBG. However, if the periodic potential has an imperfection, e.g. a vacancy which break the translational symmetry, in addition to the energy band structure, an energy gap appears between the ground state and the first excited state of energies, which promotes BEC in any dimension $d>0$  \cite{Guillermo}. For ideal Bose gases in 1D, 2D and 3D perfect and imperfect periodic structures, its BEC critical temperature in addition to some thermodynamic properties, has been reported \cite{Juan-TM,Emilio-TL,Paty}.

Although BEC of IBG is fundamental to understanding many macroscopic quantum phenomena like superfluidity \cite{Halperin,Stringari}, this is not enough, since one needs to also consider interactions between bosons as well as other features of quantum systems, e.g., point defects in solid helium-4 are necessary to understand its supersolidity, which could be caused by the coherent motion of its vacancies as suggested by Andreev \cite{Andreev} and Chester \cite{Chester}.

It is well known that the ground state energy for a {\it free} interacting one dimensional Bose gas has been calculated analytically \cite{LL} being a reference for other ground state energy calculations of more complex one-dimensional bosonic systems for which only approximations and/or computer simulation are available \cite{Optical Lattice1,Optical Lattice2}.
Furthermore, the possibility of trapping atoms in optical lattices has increased the interest in theoretically obtaining the properties of bosons within perfect periodic structures with the motivation of being able to compare them with those measured experimentally \cite{Omar, Astrakharchik}.

 In this context, Seaman et al. \cite{Carl} analyzed in detail the nonlinear band structure of a weakly interacting Bose Gas in a Dirac Comb potential within the framework of the nonlinear Schrödinger (Gross–Pitaevskii) equation. They obtained solutions for Bloch states in both repulsive and attractive interaction regimes with the appearance of swallowtails in the energy bands. %These loop structures arise from the nonlinear Gross-Pitaevskii equation which indicates the existence of multiple stationary flow states for the same quasimomentum. In physical terms, this means that the condensate can sustain persistent currents through the periodic potential, which is a hallmark of superfluidity. 
Moreover, they mapped the stability regions of these states and showed that the Dirac Comb model qualitatively reproduces the behavior of sinusoidal optical lattices while providing the advantage of more tractable analytical solutions.

Here, for a weaker interacting Bose gas trapped by a one-dimensional imperfect artificial crystal, in the framework of  the mean field approximation we study the effect of the presence of point defects (vacancies) in the artificial crystal generated with an external Dirac Comb potential, on its ground state properties.

%Furthermore, since the Chester \cite{Chester} and Andreev \cite{Andreev} suggestions that the supersolidity of helium-4 solid For example, using quantum Monte Carlo simulations, it has been studied vacancy effect on the thermodynamic properties of different quantum gases such as the energy of vacancy creation in solid $^4$He \cite{Jordi}.
%
%and an anomaly in the specific heat that is reminiscent of a second-order phase transition in these systems \cite{Jordi2} has even been found for a weakly interacting one-dimensional boson gas. 

% For the interacting 1D Bose gas trapped by an external periodic potential with vacancies ...  \cite{Jordi}. 
%
%
%In the mean field approximation, the weakly interacting boson gas within periodic structures has
%been studied and the corresponding Gross-Pitaevskii \cite{Kengne} equation has been solved
%analytically, as an example of this we also have the weakly interacting gas within multi-bars
%\cite{Omar}. \\
%
%

In Sec. II we describe our problem and the formalism to solve it. In Sec. III we solve the corresponding Gross-Pitaevskii equation using the imaginary time method. In Sec. IV the way in which the system deltas were modeled and the type of systems to be analyzed were described. In Sec. V we give the energy of the ground state, the chemical potential and the probability density function of the analyzed systems. In Sec. VI we present our conclusions.

% for a system of many quantum particles to present superfluidity it
%is necessary for the particles to interact with each other,

\section{One-dimensional Gross-Pitaevskii equation}
We solved several systems composed of a gas of weakly interacting bosons trapped by a one-dimensional artificial crystal where different percentages of the total deltas are removed randomly. 

The one-dimensional perfect finite crystal is modeled by applying an external Dirac comb potential to the Bose gas, defined as \cite{KP} 
\begin{equation}
  \label{eq:dirac-comb-potential}
  V(x)=v_0 \sum_{j=-M/2}^{M/2}\delta(x-ja)
\end{equation}
where $v_0$ is the delta strength, $a$ is the potential period, the position of each delta is $ja$ with $j \in [-M/2, M/2]$ and $M$ a positive even integer, such that the crystal length is $L= M a$.

Since we think of weakly interactions, we take mean-field theory and the system dynamics is described by the Gross-Pitaevskii equation (GPE)~\cite{Stringari}
\begin{equation}
  \label{eq:gpe-time-dependent}
  i \hbar \frac{\partial}{\partial t} \Psi(x, t) = \hat{H}_{GP}^{(t)} \Psi(x, t)
\end{equation}
where ${\hat{H}}_{GP}^{(t)}$ is the time-dependent Gross–Pitaevskii Hamiltonian
\begin{equation}
  \label{eq:gp-equation}
  {\hat{H}}_{GP}^{(t)} =- \frac{\hbar^2}{2m} \frac{\partial^2}{\partial x^2} + \extpotential(x) + g |\Psi(x, t)|^2,
\end{equation}
and $\Psi(x, t)$ is the wave function of the boson system, $m$ is the mass of each boson, $g$ is a
parameter that measures the strength of the interaction between bosons, and $\extpotential(x)$ is
the external potential,~\eqref{eq:dirac-comb-potential}. The system size is $L$ and~\eqref{eq:gpe-time-dependent} is subject to periodic boundary conditions, i.e., $\Psi(-L/2, t)= \Psi(L/2, t)$.

The system has two important, conserved properties: its energy,
\begin{align}
  E(\Psi(\cdot, t)) & = \int_L \left( \frac{\hbar^2}{2m} \left|\frac{\partial \Psi(x, t)}{\partial x} \right|^2 + \right. \nonumber \\
                    & \left. \extpotential(x)\left|\Psi(x, t)\right|^2 + \frac{g}{2}\left|\Psi(x, t) \right|^4 \right) \; dx,
  \label{eq:system-energy}
\end{align}
and the number of particles $N$,
\begin{equation}
  \label{eq:norm-cond-wf}
  N(\Psi(\cdot, t)) = \int_L \left| \Psi(x, t) \right|^2 \; dx,
\end{equation}
where $\Psi(\cdot, t)$ indicates that $E$ and $N$ are time-dependent functionals after the positional integration of the wave function. % and other quantities.
Equation~\eqref{eq:norm-cond-wf} is also referred to as the normalization condition.

In the present work, we are interested in the quantum stationary states, which evolve as
\begin{equation}
  \label{eq:stationary-states}
  \Psi(x, t) = \Phi(x) e^{i \mu t /\hbar},
\end{equation}
where $\mu$ is the chemical potential of the system. For these states, the time-dependent
GPE~\eqref{eq:gpe-time-dependent} becomes the stationary Gross-Pitaevskii equation,
\begin{equation}
  \label{eq:GPmu}
  \hat{H}_{GP} \Phi(x) = \mu \Phi(x),
\end{equation}
where
\begin{equation}
  \label{eq:GP1D}
  \hat{H}_{GP} = -\frac{\hbar^2}{2m} \frac{\partial^2}{\partial x^2} + \extpotential(x)+ g \left| \Phi(x) \right|^2.
\end{equation}
%
% where $g_{1D} = {-2\hbar^2}/{ma_{1D}}$ and $a_{1D} = a_{\perp}^2/a_s$
From~\eqref{eq:GPmu} and \eqref{eq:GP1D}, we can easily relate the chemical potential and the energy per particle as
\begin{equation}
  \mu = \frac{E(\Phi(\cdot))}{N} + \frac{g}{2N} \int_L \left|\Phi(x) \right|^4 \; dx
\end{equation}
where $E(\Phi(\cdot))$ is
\begin{align}
  E(\Psi(\cdot)) & = \int_L \left( \frac{\hbar^2}{2m} \left|\frac{\partial \Psi(x)}{\partial x} \right|^2 + \right. \nonumber \\
                    & \left. \extpotential(x)\left|\Psi(x)\right|^2 + \frac{g}{2}\left|\Psi(x) \right|^4 \right) \; dx,
  \label{eq:system-energy2}
\end{align}
and the number of particles $N$,
\begin{equation}
  \label{eq:norm-cond-wf2}
  N(\Psi(\cdot)) = \int_L \left| \Psi(x) \right|^2 \; dx,
\end{equation}

%copiar la ecuación 4 y 5 quitando la dependencia del tiempo.
For a non-interacting system, it follows the chemical potential is equal to the ground state energy per
particle, as expected.

%Que es $E(\Phi(\cdot)$

\subsection{Dimensionless Gross-Pitaevskii Equation}

To obtain the ground state physical properties of our system, we have to write the GPE in a
dimensionless way and solve it numerically. We start by defining units for the length, time, and
energy. Let us take inspiration from the quantum harmonic oscillator and define a
\emph{characteristic length}, $\charlen$, and a \emph{characteristic angular frequency}, $\omegac$,
related by the equation
\begin{equation}
  \label{eq:dimlesslen}
  \charlen = \sqrt{\frac{\hbar}{m \omegac}}.
\end{equation}
Also, let us assume the system has an average (linear) density $n$ and $\chardensity \equiv N / \charlen$. These definitions, together with~\eqref{eq:dimlesslen} and
the following definitions,
\begin{subequations}
  \begin{equation}
    x = \charlen x', \quad t = t' / \omegac, \quad \Psi(x, t) = \sqrt{\chardensity} \, \Psi'(x', t'),
  \end{equation}
  \begin{equation}
    \gpeham^{(t)} = \gpeham'^{(t)} \hbar \omegac, \quad \extpotential = \extpotential'  \hbar \omegac, \quad g = g' \hbar \omegac / n
  \end{equation}
\end{subequations}
that relate the length, time, energy, and wave-function expressed in physical units (unprimed
symbols) to their corresponding dimensionless counterparts (primed symbols). Then we can rewrite the
GPE~\eqref{eq:gpe-time-dependent} in the following way
\begin{equation}
  \label{eq:gpe-dimless}
  i\frac{\partial}{\partial t'} \Psi'(x', t') = \gpeham'^{(t)} \Psi'(x', t').
\end{equation}
Accordingly, the dimensionless GPE Hamiltonian~\eqref{eq:gp-equation} is given by
\begin{equation}
  \label{eq:gpe-dimless-hamiltonian}
  \gpeham'^{(t)} = -\frac{1}{2} \frac{\partial^2}{\partial x'^2} + \extpotential'(x') + g'\frac{\chardensity}{n} \left| \Psi'(x', t') \right|^2.
\end{equation}
Analogously, the system energy per particle, $E/N$, and its dimensionless counterpart,
$E' = E / N \hbar \omegac$, yield the following result,
\begin{align}
  \nonumber
  E'(\NormPsi(\cdot, t')) & =                 \int_{L'} \left( \frac{1}{2} \left|\frac{\partial \Psi'(x', t')}{\partial x'} \right|^2 + \right.      \\
  \label{eq:system-energy-dimless}
  \extpotential'(x')      & \left. \left| \Psi(x', t') \right|^2 + \frac{g'}{2} \frac{\chardensity}{n} \left|\Psi'(x', t') \right|^4 \right) \; dx'.
\end{align}
Finally, the normalization condition becomes
\begin{equation}
  \label{eq:norm-condition-wf-dimless}
  \int_{L'} \left| \Psi'(x', t') \right|^2 \; dx' = 1.
\end{equation}
Regarding the stationary states, we have
\begin{equation}
  \Psi'(x', t') = \Phi'(x') e^{i \mu' t'},
\end{equation}
where $\mu = \mu' \hbar \omegac$ and $\Phi(x) = \sqrt{\chardensity} \Phi'(x')$. Then, the stationary
Gross-Pitaevskii equation becomes
\begin{equation}
  \label{eq:gpe-stationary-dimless}
  \gpeham' \Phi'(x') = \mu' \Phi'(x')
\end{equation}
where $\gpeham'=\gpeham/ \hbar \omegac$.
Accordingly,
\begin{equation}
  \mu' = E'(\Phi'(\cdot)) + \frac{g'}{2} \frac{\chardensity}{n} \int_L' \left|\Phi'(x') \right|^4 \; dx'.
\end{equation}

From now on, having established the dimensionless form of the GPE and other important quantities, to
simplify the mathematics, we will assume we are always working with dimensionless quantities and
drop the primes.

\section{Mathematical Approach to Determine the Ground-State}

% The ground state of our system satisfies the equation~\eqref{eq:gp-equation}. At the same time, it
% minimizes eq.~\eqref{eq:system-energy-dimless} under the normalization
% condition~\eqref{eq:norm-condition-wf-dimless}. To find the ground-state, we employ a \emph{Gradient
% Flow with Discrete Normalization} (GDFN) \cite{Bao} method. This procedure is a gradient-flow algorithm  that
% forces the normalization condition is also commonly known as the
% \emph{imaginary time} method since the equation we solve is equal to the GPE written in imaginary
% time.

The ground state has a wave function that minimizes the system
energy~\eqref{eq:system-energy-dimless} subject to the constraint imposed by the normalization
condition~\eqref{eq:norm-condition-wf-dimless}, i.e., we must solve a constrained
optimization problem.

% The energy of the system is the expected value of the Hamiltonian, and we can write it as
% %
% \begin{align}
%   \nonumber
%   E & \equiv E(\Psi)                                                                                                  \\
%     & = \left\langle \Psi \left| \hat{H} \Psi \right. \right\rangle                                                   \\
%     & = \left\langle \Psi \left| \gpeham \Psi \right. \right\rangle - \frac{g}{2} \int \left| \Psi(z, t) \right|^4 dz
% \end{align}

We will use a Gradient Flow approach to minimize the system free energy, $F = E - \mu N$, given the
system dynamics is constrained by the wave function normalization
condition~\eqref{eq:norm-condition-wf-dimless}. As its name suggests, the method depends on the
gradient of the Free energy to obtain the ground state. However, we must recall that $F$ is a
functional of the wave function $\Psi(x, t)$, its derivative, $\partial_x \Psi(x, t)$, and the
complex conjugate of both, $\Psi^*(x, t)$, and $\partial_x \Psi^*(x, t)$, respectively.
Consequently, the \emph{functional derivative} of the Free energy $F$ plays the role of the gradient
in the associated gradient flow equation of our optimization problem,
\begin{equation}
  \label{eq:functional-gradient-flow}
  % \frac{\partial \Psi(x, \tau)}{\partial \tau} = -\frac{\delta}{\delta \Psi^*(x, \tau)} F(\Psi(\cdot, \tau), \partial \Psi(\cdot, \tau)),
  \frac{\partial \Psi}{\partial \tau} = -\frac{\delta}{\delta {\Psi}^*} F \left(\Psi(\cdot), \partial \Psi(\cdot), {\Psi}^*(\cdot), \partial {\Psi}^*(\cdot) \right),
\end{equation}
where
\begin{align}
  \nonumber
  % \hspace{-0.3cm} F(\Psi(\cdot, \tau), \partial \Psi(\cdot, \tau)) = \int_{L} f(x, \Psi(x, \tau), \partial_x \Psi(x, \tau)) \; dx \;
   & F \left(\Psi(\cdot), \partial \Psi(\cdot), {\Psi}^*(\cdot), \partial {\Psi}^*(\cdot) \right) =    \\
  \label{eq:functional-free-energy}
  %  & \qquad \qquad \int_{L} f(x, \Psi(x, \tau), \partial_x \Psi(x, \tau)) \; dx \;
   & \qquad \qquad \int_{L} f \left(\Psi, \partial_x \Psi, {\Psi}^*, \partial_x {\Psi}^* \right) \; dx
\end{align}
%- \\
%  & \qquad \qquad \mu \int_{L} \left|\Psi(x, \tau) \right|^2 \; dx
%
and
\begin{align}
  \nonumber
   & f \left(\Psi, \partial_x \Psi, {\Psi}^*, \partial_x {\Psi}^* \right)  = \frac{1}{2} \frac{\partial {\Psi^*}}{\partial x} \frac{\partial \Psi}{\partial x} + \\
  %  & f(x, \Psi(x, \tau), \partial_x \Psi(x, \tau)) = \frac{1}{2} \left|\frac{\partial \Psi(x, \tau)}{\partial x} \right|^2 +    \\
  \label{eq:functional-free-energy-integrand}
   & \qquad \qquad (\extpotential(x)- \mu)  \Psi^* \Psi + \frac{g}{2} \frac{\chardensity}{n} \left(\Psi^*\right)^2 \left(\Psi\right)^2.
  %  & \qquad (\extpotential(x)- \mu)  \left| \Psi(x, \tau) \right|^2 + \frac{g}{2} \frac{\chardensity}{n} \left|\Psi(x, \tau) \right|^4.
\end{align}
For the sake of conciseness, we omitted the explicit functional dependence $\Psi(x, \tau)$ of the
wave function in eqs.~\eqref{eq:functional-gradient-flow}, \eqref{eq:functional-free-energy}, and
\eqref{eq:functional-free-energy-integrand}. As in the eq.~\eqref{eq:system-energy}, the $(\cdot)$
notation indicates that an integration over the positional variable is present in
eq.~\eqref{eq:functional-gradient-flow}. Accordingly, the Free energy depends only on $\tau$ because
it inherits the wave function dependence of $\tau$, while the dependence on $x$ vanishes due to the
positional integration.

Equation~\eqref{eq:functional-gradient-flow} models a process (the
optimization) where the wave function $\Psi(x, \tau)$ evolves from its initial value, $\Psi(x, \tau
  = 0)$, while being subject to a set of boundary conditions specific to the system being studied, for
example, Dirichlet or Neumann, towards the ground state following the direction of the steepest
descent of the Free energy, i.e., the negative of its gradient. Note we have used the variable
$\tau$ instead of the physical time $t$ because, in the context of optimization using the gradient
flow method, $\tau$ is a parameter that defines how the initial condition evolves towards the ground
state. In some sense, it can be thought as a time, although not a physical one, despite the
resemblance between both.

The machinery of calculus of variations states the functional derivative of the Free energy is
\begin{equation}
  \frac{\delta}{\delta \Psi^*} F =\frac{\partial f}{\partial \Psi^*} - \frac{\partial}{\partial x} \frac{\partial f}{\partial (\partial_x \Psi^*)}.
\end{equation}
Therefore, substituting equation %
\begin{align}
  \nonumber
   & \frac{\delta}{\delta \Psi^*(x, \tau)} F(\Psi(\cdot, \tau), \partial \Psi(\cdot, \tau)) =                                                                           \\
  \nonumber
   & \qquad \left( -\frac{1}{2} \frac{\partial^2}{\partial x^2} + \extpotential(x) + g \frac{\chardensity}{n} \left| \Psi(x, \tau) \right|^2 \right) \Psi(x, \tau) \; - \\
   & \qquad \qquad \mu \Psi(x, \tau).
\end{align}
% como llegamos a la ecuación (23)?
In this way, the gradient flow equation~\eqref{eq:functional-gradient-flow} becomes
\begin{align}
  \nonumber
   & \frac{\partial \Psi(x, \tau)}{\partial \tau} =                                                                                                                    \\
  \nonumber
   & \qquad \left( \frac{1}{2} \frac{\partial^2}{\partial x^2} - \extpotential(x) - g \frac{\chardensity}{n} \left| \Psi(x, \tau) \right|^2 \right) \Psi(x, \tau) \; + \\
  \label{eq:bec-gradient-flow}
   & \qquad \qquad \mu \Psi(x, \tau).
\end{align}
The above equation~\eqref{eq:bec-gradient-flow} is exactly the same as~\eqref{eq:gpe-dimless}, plus
a term containing the chemical potential, expressed in imaginary time by means of a Wick rotation
\cite{Succi}, $\tau = i t$, so we go from the physical time to the imaginary
time. The term containing the chemical potential ensures the wave function remains normalized, i.e.,
we can prove, after a few steps, that
\begin{align}
  \frac{\partial}{\partial \tau} \int_{L} {\Psi(x, \tau)}^*\Psi(x, \tau) \; dx = 0.
\end{align}

\subsection{Gradient Flow with Discrete Normalization}
% Que tan cerca estamos del estado base-comentario para Omar
The equation~\eqref{eq:bec-gradient-flow} models the evolution of an initial wave function towards that of
the ground state keeping the wave function normalization. In principle, if we solve it by
means of any analytical or numerical method satisfying the initial and boundary conditions, then we
obtain the ground state. However, the chemical potential appears in~\eqref{eq:bec-gradient-flow}, so
to apply it we would have to first solve the Eq.~\eqref{eq:gpe-stationary-dimless} subject
to~\eqref{eq:norm-condition-wf-dimless}, which describes a tautological procedure. To solve the above
difficulty, we use a numerical two-step scheme to evolve the initial wave function and fulfill the
normalization condition referred to as \emph{Gradient-Flow with Discrete
  Normalization} (GFDN)~\cite{bib:bao-2004}.

First, we define an (imaginary) time sequence $0 = \tau_0 < \tau_1 < \tau_2 < \cdots < \tau_n <
  \cdots$ with a given time-step $\Delta \tau_n = (\tau_{n+1} - \tau_n) > 0$, and an initial
approximation $\Psi(x, \tau = \tau_0 = 0)$. Then, during the interval $\tau_n < \tau < \tau_{n+1}$,
we evolve the initial wave function, $\Psi(x, \tau_n)$, following the unconstrained gradient flow
equation, \eqref{eq:bec-gradient-flow} with $\mu=0$,
\begin{align}
  \nonumber
   & \frac{\partial \Psi(x, \tau)}{\partial \tau} =                                                                                                                \\
  \label{eq:bec-gradient-flow-no-constraint}
   & \qquad \left( \frac{1}{2} \frac{\partial^2}{\partial x^2} - \extpotential(x) - g \frac{\chardensity}{n} \left| \Psi(x, \tau) \right|^2 \right) \Psi(x, \tau),
\end{align}
subject to some boundary conditions. At the end of the evolution, we get a new approximation of the
wave function, $\Psi(x, \tau_{n+1})$. However, this new approximation no longer satisfies the
normalization condition, so we force it to satisfy it
\begin{equation}
  \label{eq:discrete-normalization}
  \Psi(x, \tau_{n+1}) \rightarrow \frac{\Psi(x, \tau_{n+1})}{\lVert \Psi(x, \tau_{n+1}) \rVert},
\end{equation}
where $\lVert \Psi(x, \tau_{n}) \rVert = \int_{L} \left|\Psi(x, \tau_n) \right|^2 dx$ for all $n=0, 1, ...$

From the numerical perspective, at a given time interval, first, we solve the gradient flow
equation~\eqref{eq:bec-gradient-flow-no-constraint} by any method at our hand. Then, at the end of
the time step, we normalize the wave function by force. The above
equation~\eqref{eq:bec-gradient-flow-no-constraint} is equal to time-dependent
GPE~\eqref{eq:gpe-dimless} expressed in the imaginary time $\tau = i t$. For this reason, the GFDN
method is also commonly known as the imaginary time method. At this point, our main challenge is
solving the equation~\eqref{eq:bec-gradient-flow-no-constraint}. Multiple methods to solve this kind
of partial differential equation have been used in the past \cite{Adhikari}.

In this work, we use a Backward Euler finite-difference discretization in time, and a Discrete
Fourier Transform (DFT) based pseudo-spectral approach to solve the spatial part, as described
in~\cite[]{Bao}. At each step of the time sequence $\tau_n$, we approximate
\begin{align}
  \nonumber
  \left. \frac{\partial \Psi(x, \tau)}{\partial \tau} \right|_{\tau = \tau_n} & \approx \frac{\Psi(x, \tau_{n+1}) - \Psi(x, \tau_n)}{\Delta \tau} \\
  \label{eq:backward-euler-gradient-flow-approx}
                                                                              & = \mathcal{F}(x, \tau_{n+1}),
\end{align}
where $\mathcal{F}(x, \tau_{n+1})$ is the r.h.s.\ of~\eqref{eq:bec-gradient-flow-no-constraint}
evaluated at $\tau = \tau_{n+1}$. The Backward Euler is an implicit method, because the r.h.s.\
of~\eqref{eq:backward-euler-gradient-flow-approx} is evaluated at $\tau_{n+1}$, the start of the
next time step, instead of the current one.

In the pseudo-spectral approach, we express the wave function as a finite, linear combination of
plane waves of wave vectors $q_k$ and amplitudes $\planeWaveAmplitude(\tau)$,
\begin{equation}
  \label{eq:ground-state-fourier-series}
  \Psi(x, \tau) = \sum_{k} \planeWaveAmplitude(\tau) e^{i q_k (x - x_0)},
\end{equation}
% esta normalizada la ecuación (28)
where the domain $L$ ranges from $x_0$ to $x_f$, so $x_f - x_0 = L > 0$. By using the above
expression~\eqref{eq:ground-state-fourier-series} we are implicitly fulfilling periodic boundary
conditions on the wave function as prescribed by the Dirac-comb
potential~\eqref{eq:dirac-comb-potential}, i.e., $\Psi(x_0, \tau) = \Psi(x_f, \tau)$ with $x_0=-L/2$ and $x_f=L/2$.

In the DFT, pseudo-spectral approach, we discretize the domain in a set of $M-1$ points $x_j = x_0 +
  j \Delta_x$, $j = 0, 1, \ldots, M-1$, i.e., we are partitioning the domain $L$ into $M$
equally-sized segments of size $\Delta_x = (x_f - x_0) / M = L / M$. Note we omit the point $x_M =
  x_f$ because of the periodic boundary conditions. Accordingly, for each point $x_j$, the wave
function fulfills
\begin{align}
  \label{eq:ground-state-fourier-series-equations}
  \psi_j(\tau) & = \sum_{k} \planeWaveAmplitude(\tau) e^{i j q_k \Delta_x / M}, \\
  \nonumber
  j, k         & = 0, 1, \ldots, M-1,
\end{align}
where $\psi_j(\tau) = \Psi(x_j, \tau)$. Since only $M-1$ of the plane waves
in~\eqref{eq:ground-state-fourier-series-equations} are linearly independent, the above set of
equations relate the $M-1$ plane wave amplitudes with the $M-1$ wave function values. The former can
be efficiently calculated by means of the Fast-Fourier Transform algorithm, a feature that
potentially makes solving the GPE by the DFT, pseudo-spectral approach very fast compared to other
methods like finite-differences.
% mejorar la redacción del siguiente parrafo
After solving the differential equation~\eqref{eq:backward-euler-gradient-flow-approx}, we have
estimates, at $\tau_{n + 1}$, of the wave function at every domain partition point $x_j$, and the
plane wave amplitudes for every wave vector $q_k$, i.e., $\psi_j(\tau_{n+1})$ and
$\planeWaveAmplitude(\tau_{n+1})$, respectively. From the above estimates, can determine the
energy~\eqref{eq:system-energy-dimless} and the chemical potential as
\begin{align}
  \nonumber
   & E^{n + 1} \approx \frac{L}{4} \sum_{k} q_k^2 \left| (\planeWaveAmplitude)^{n + 1} \right|^2 +                                                  \\
   & \qquad  \Delta_x \sum_{j} V_j \left| \psi_j^{n+1} \right|^2 + \frac{g}{2} \frac{\chardensity}{n} \Delta_x \sum_{j} \left|\psi_j^{n+1}\right|^4
\end{align}
and
\begin{equation}
  \mu^{n+1} \approx E^{n+1} + \frac{g}{2} \frac{\chardensity}{n} \Delta_x \sum_{j} \left|\psi_j^{n+1}\right|^4
\end{equation}
where $\psi_j^{n+1} = \psi(x_j, \tau_{n+1})$, ${(\tilde{\psi}_k)}^{n+1} = \planeWaveAmplitude(\tau_{n+1})$, $E^{n + 1} = E(\Psi(\tau_{n+1}))$, $\mu^{n+1} = \mu(\Psi(\tau_{n+1}))$, and $V_{j} = \extpotential(x_j)$.

\section{Random vacancies}

%\subsection{Deltas}
Before we show that the external potential only depends on the spatial grid, with this in mind we modeled the deltas raising one point of the partition periodically, breaking this periodicity we introduce the vacancies.

Different systems were analyzed, varying strength of the interaction $g$ and the vacancies. %each one had an interaction $g=0$, $1$ and $2$ respectively.

\begin{figure}[H]
  \centering
  \includegraphics[width=\linewidth]{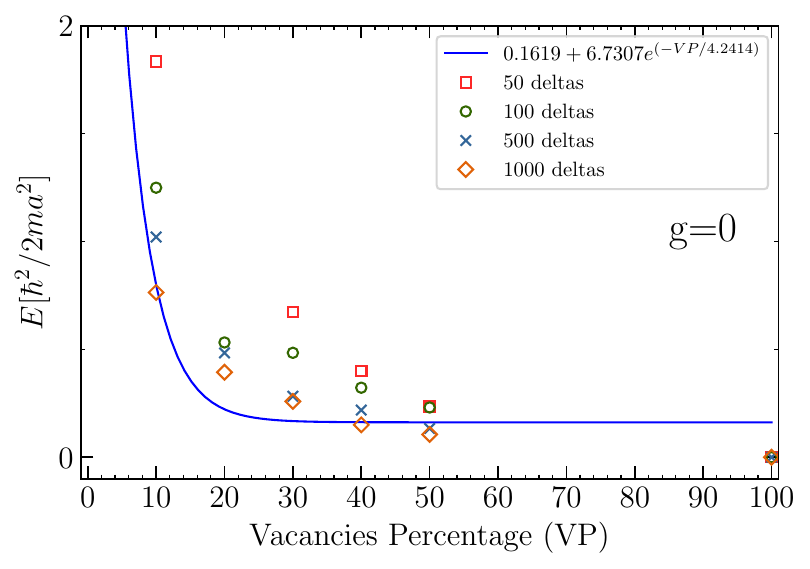}
  \caption{Ground state energy, in units of $\hbar^2/2ma^2$, for an ideal Bose gas ($g=0$) within of 50, 100, 500
    and 1000 initial deltas where we randomly removed 10\%, 20\%, 30\%, 40\% y 50\%, $P_0=10$. The full curve is an exponential fit for the 1000 delta values.}%
  \label{fig:Energiasg0}
\end{figure}

\begin{table}[H]
  \begin{center}
    \begin{tabular}{c|llll}
      \toprule
      Removed & \multicolumn{4}{c}{Initial number of deltas}                            \\
      \midrule
      \%      & 50                                           & 100    & 500    & 1000   \\ \hline
      0       & 6.8942                                       & 6.8942 & 6.8942 & 6.8942 \\
      10      & 1.8338                                       & 1.2496 & 1.0210 & 0.7638 \\
      20      & 1.0511                                       & 0.5320 & 0.4836 & 0.3943 \\
      30      & 0.6741                                       & 0.4841 & 0.2827 & 0.2591 \\
      40      & 0.3987                                       & 0.3224 & 0.2187 & 0.1508 \\
      50      & 0.2343                                       & 0.2307 & 0.1332 & 0.1056 \\
      100     & 0                                            & 0      & 0      & 0      \\
      \bottomrule
    \end{tabular}
    \caption{ Values of the ground state energy, in units of $\hbar^2/2ma^2$, for a Bose gas within of  50, 100, 500 y 1000 initial deltas, where we randomly removed 10\%, 20\%, 30\%, 40\% and 50\%, with $P_0 = 10$ and $g = 0$.}
  \label{table:tabla1}  
  \end{center}
\end{table}

\begin{figure}[H]
  \centering
  \includegraphics[width=\linewidth]{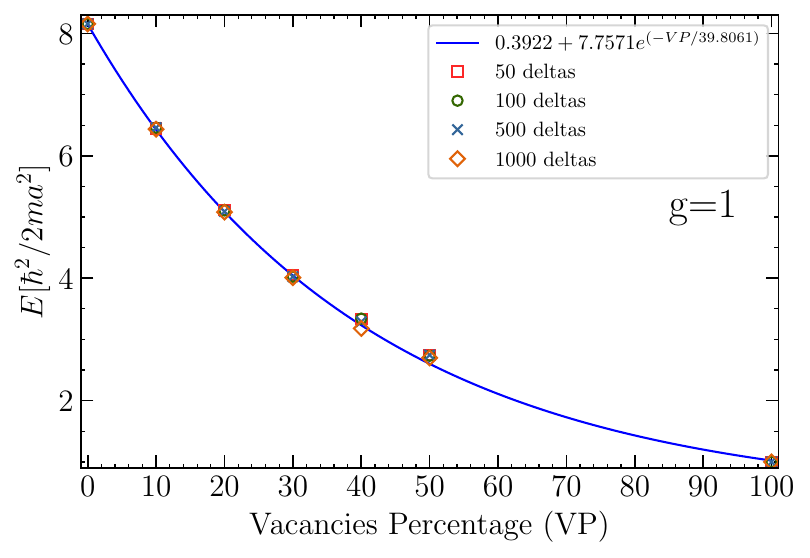}
  \caption{Ground state energy, in units of $\hbar^2/2ma^2$, for a Bose gas within of 50, 100, 500 and 1000 total initial deltas where we randomly removed 10\%, 20\%, 30\%, 40\% and 50\%, with $P_0 = 10$ and $g = 1$.We did an exponential fit on the 1000 delta values, this is shown in the curve.}%
  \label{fig:Energiasg1}
\end{figure}

\begin{table}[H]
  \begin{center}
    \begin{tabular}{c|llll}
      \toprule
      Removed & \multicolumn{4}{c}{Initial number of deltas}                            \\
      \midrule
      \%      & 50                                           & 100    & 500    & 1000   \\ \hline
      0       & 8.1543                                       & 8.1543 & 8.1543 & 8.1543 \\
      10      & 6.4482                                       & 6.4598 & 6.4577 & 6.4637 \\
      20      & 5.1183                                       & 5.1084 & 5.0901 & 5.0623 \\
      30      & 4.0520                                       & 4.0264 & 4.0252 & 4.0268 \\
      40      & 3.3315                                       & 3.3395 & 3.2895 & 3.2562 \\
      50      & 2.7428                                       & 2.7335 & 2.7460 & 2.7332 \\
      100     & 0.9999                                       & 0.9999 & 0.9999 & 0.9999 \\
      \bottomrule
    \end{tabular}
    \caption{Values of the ground state energy, in units of $\hbar^2/2ma^2$, for a Bose gas within of  50, 100, 500 y 1000 total initial deltas, where we removed  10\%, 20\%, 30\%, 40\% and 50\%, with $P_0 = 10$ and $g=1$.}
 \label{table:tabla2}  
  \end{center}%
\end{table}

\begin{figure}[H]
  \centering
  \includegraphics[width=\linewidth]{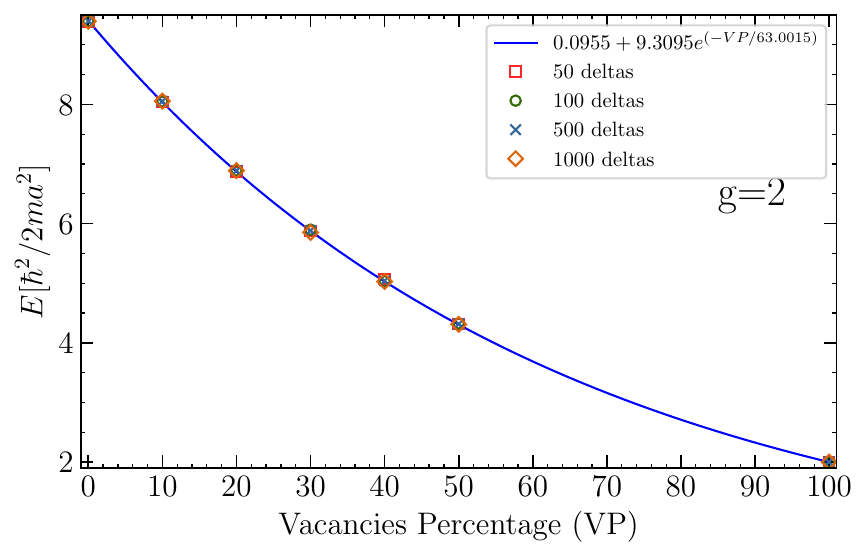}
  \caption{Ground state energy for systems with 50, 100, 500 and 1000 initial deltas where we removed
    10\%, 20\%, 30\%, 40\% and 50\%, with $P_0=10$ and $g=2$.We did an exponential fit on the 1000 delta values, this is shown in the curve.}%
  \label{fig:Energiasg2}
\end{figure}

\begin{table}[H]
  \begin{center}
    {\small    \begin{tabular}{c|llll}
        \toprule
        Removed & \multicolumn{4}{c}{Number of initial deltas}                              \\
        \midrule
        \%      & 50                                           & 100    & 500     & 1000    \\ \hline
        0       & 9.3942                                       & 9.3942 & 9.3942  & 9.3942  \\
        10      & 8.0428                                       & 8.0521 & 8.0520  & 8.0542  \\
        20      & 6.8778                                       & 6.8913 & 6.8900  & 6.8881  \\
        30      & 5.8778                                       & 5.8987 & 5.8852  & 5.8832  \\
        40      & 5.0747                                       & 5.0327 & 5.03915 & 5.03910 \\
        50      & 4.3194                                       & 4.3252 & 4.3156  & 4.3292  \\
        100     & 2                                            & 2      & 2       & 2       \\
        \bottomrule
      \end{tabular}
    } \caption{Ground state energy (GSE) values in units of $\hbar^2/2ma^2$, for a Bose gas within of  50, 100, 500
      and 1000 total initial deltas, where we removed  10\%, 20\%, 30\%, 40\% and 50\%, with $P_0=10$ and $g=2$.}
        \label{table:tabla3}
  \end{center}%
\end{table}
\vspace{-1.2cm}
\begin{figure}[H]
  %\centering
  \includegraphics[height=6cm, width=8cm]{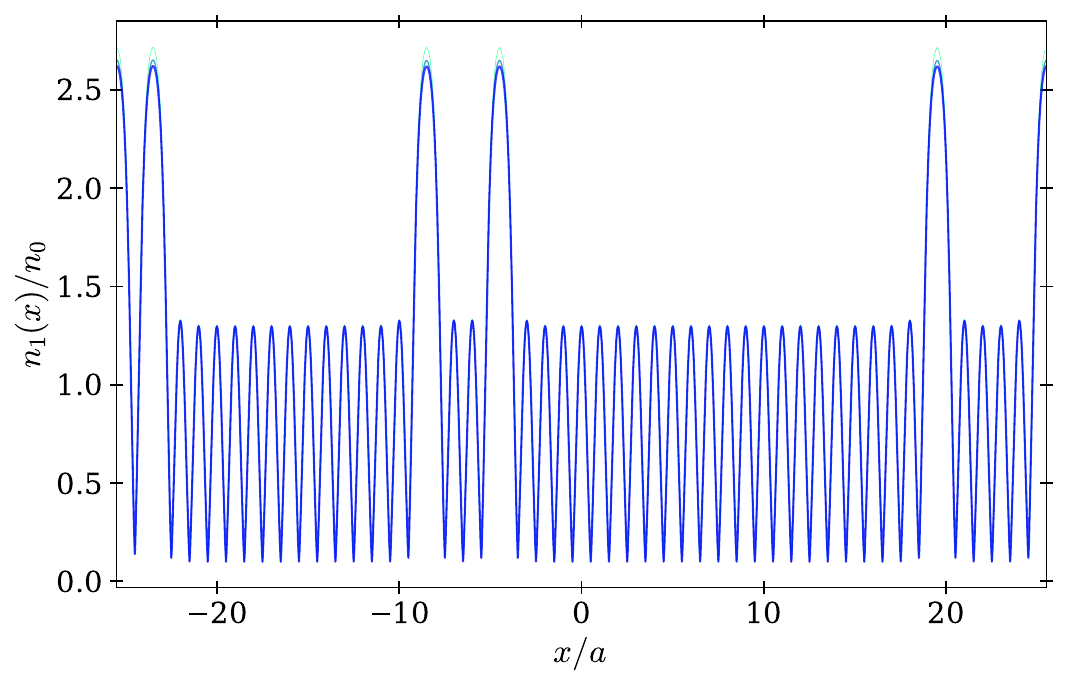}
  \vspace{-0.5cm}
  \caption{Probability density function of the ground state of a system with 50
    deltas where 10\% of random deltas were removed, $P_0=10$ and $g=2$.}%
  \label{fig:FunOnda1}
\end{figure}

\begin{figure}[H]
  %  \centering
  \begin{center}
    \includegraphics[height=7.5cm, width=8.0cm]{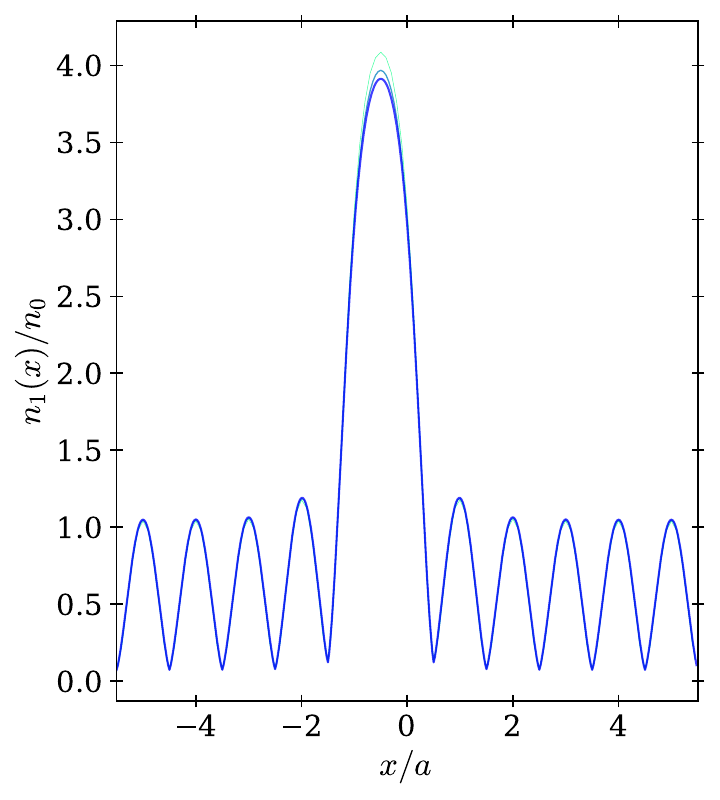}
  \end{center}
  \vspace{-0.5cm}
  \caption{Probability density function of a system with 11 deltas minus the central one, $P_0=10$
    and $g=1$}%
  \label{fig:FunOnda2}
\end{figure}

In Figs. \ref{fig:Energiasg0}, \ref{fig:Energiasg1} and \ref{fig:Energiasg2} we show the GSE %can see the results of having a certain percentage of vacancies
for systems with 50, 100, 500 and 1000 initial deltas, for several percentage of vacancies, and for $g=0, 1 and 2$ respectively. Each point corresponds to multiple samples where a given percentage of deltas was randomly removed. The
generation of new samples was stopped when the error percentage in the ground state energy was less
than 2\%, with the exception of the cases where all the deltas were removed and where none were
removed. We show that the ground state energy decreases from the perfect case (all the deltas) to the free case, the form that the decreases fit with an exponential for the tree cases, when the interaction are on the ground state energy tends to group in one point  regardless of the total deltas of the system, we show this values in the tables \ref{table:tabla1}, \ref{table:tabla2} and \ref{table:tabla3} respectively.  In the tree cases, the GSE goes to the interaction value, thats makes sense because the frontier conditions are periodic. In Figs. \ref{fig:FunOnda1} and \ref{fig:FunOnda2} we show the probability density functions as an example of the behavior of the position of particles, we notice that there is a greater probability of finding particles in positions where the deltas are removed, i.e. vacancies act as an attractive centers locating the particles. 

%again this makes sense because due to the interactions they group together more strongly when there is no delta between them.
% each of these figures has a variation in the value of the
%interaction $g$, in the respective tables we have the numerical values.
%

%\vspace{10cm}

%\vspace{10cm}

% Consistencia en puntos y espacios

\begin{figure}[H]
  %  \centering
  \includegraphics[width=\linewidth]{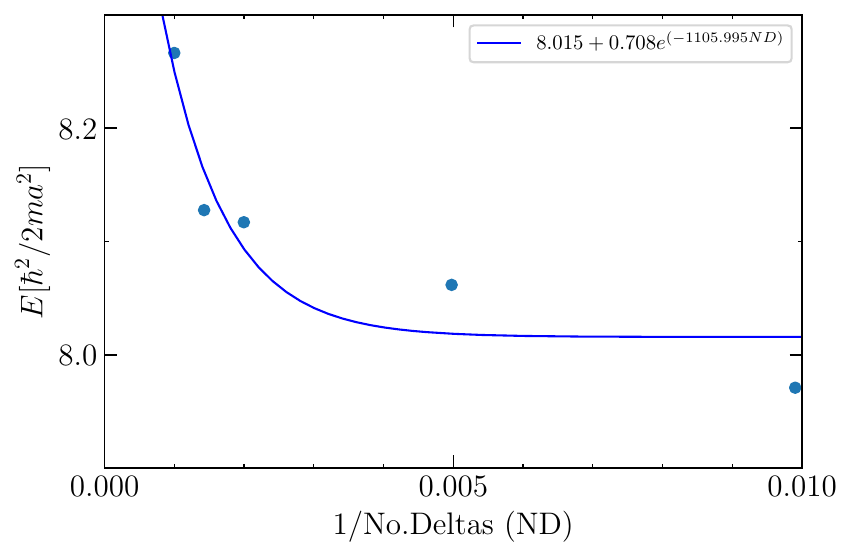}
  \caption{%Calculation of the ground state energy when the system size goes to infinity using a
    Extrapolation of the ground state energy to its thermodynamic limit value as the size system are increasing keeping its particle density constant. In the infinity limit the GSE value is $8{\mbox .}724 \ (\hbar^2/2ma^2) $ with $P_0=10$ and $g=1$.}
  \label{fig:regresion}
\end{figure}

\begin{table}[H]
  \begin{center}
    \begin{tabular}{l|lllll}
      %\toprule
                                  & \multicolumn{5}{c}{}                                  \\ \hline
                    \diagbox{System}{Values of $g$}       & 0                                 & 0.1   & 0.5   & 1     & 2      \\
      \midrule
      Free: all deltas removed    & 0                                 & 0.1   & 0.500   & 0.999 & 2.000  \\
      Perfect: zero delta removed & 6.894                             & 7.021 & 7.527   & 8.154 & 9.394  \\
      One delta removed           & 2.238                             & 6.925 & 7.499 & 8.135 & 9.379  \\
      Two delta removed           & 2.238                             & 6.829 & 7.472 & 8.117 & 9.365  \\ \hline
      Infinite: one delta removed & 2.238                             & 7.134 & 7.983 & 8.742 & 10.325 \\
      \bottomrule
    \end{tabular}
    \caption{Ground state energy (GSE) values in units of $\hbar^2/2ma^2$, for systems with 1001 total initial deltas and several values of $g$. Last row gives the GSE for a Bose gas within a infinite Dirac comb without a delta. }
      \label{table:tabla4}
  \end{center}
\end{table}
%}
%\section{Chemical potential}
%{\tiny
\begin{table}[H]
  \begin{center}
    \begin{tabular}{llllll}
      \toprule
      \diagbox{System}{Values of $g$}                        & 0     & 0.1   & 0.5   & 1     & 2      \\
      \midrule
      Free                      & 0     & 0.2   & 1.0   & 2.0   & 4.0    \\
      Perfect                   & 6.894 & 7.272 & 8.282 & 9.544 & 12.170 \\
      One delta removed         & 2.238 & 7.230 & 8.272 & 9.534 & 12.150 \\
      Two delta removed         & 2.238 & 7.210 & 8.266 & 9.522 & 11.992 \\
      One delta removed (infty) & 2.238 & 7.673 & 8.529 & 9.870 & 13.928 \\
      \bottomrule
    \end{tabular}
    \caption{Chemical potential values in units of $\hbar^2/2ma^2$ for a systems with  1001 total initial deltas.}
  \label{table:tabla5} 
  \end{center}
\end{table}
%}
%\section{Lieb-Liniger}

\begin{figure}[H]
  %	\centering
  \includegraphics[width=\linewidth]{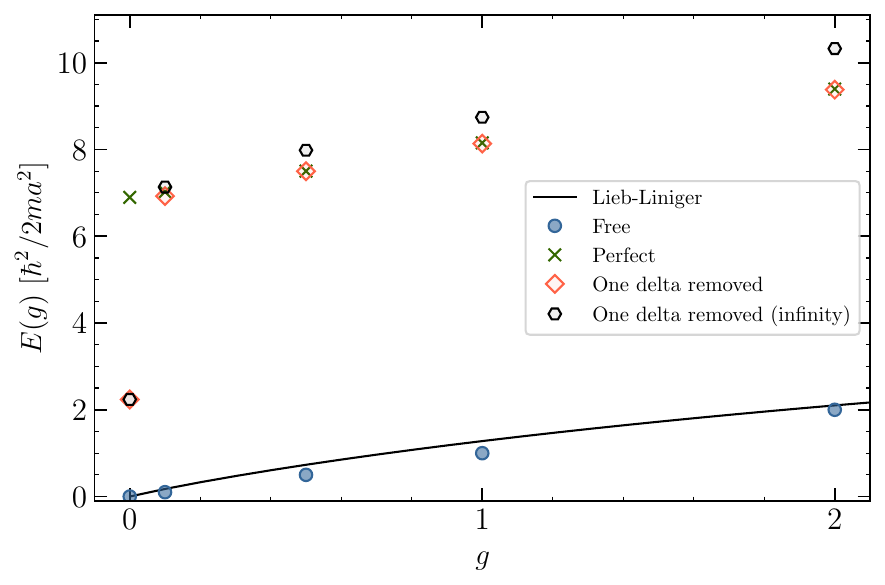}
  \caption{ Ground state energies $E(g)$ vs. $g = \gamma$ for several systems: a Lieb-Liniger Bose gas (solid line), a free weakly interacting Bose gas (filled circle), a Bose gas within a perfect 1D crystal (cross), the same Bose gas when one delta is removed (diamond) and the extrapolation for an infinite system (hexagon)}%
  \label{fig:Lieb}
\end{figure}

\begin{figure}[H]
  %	\centering
  \includegraphics[height=7.0cm, width=8.0cm]{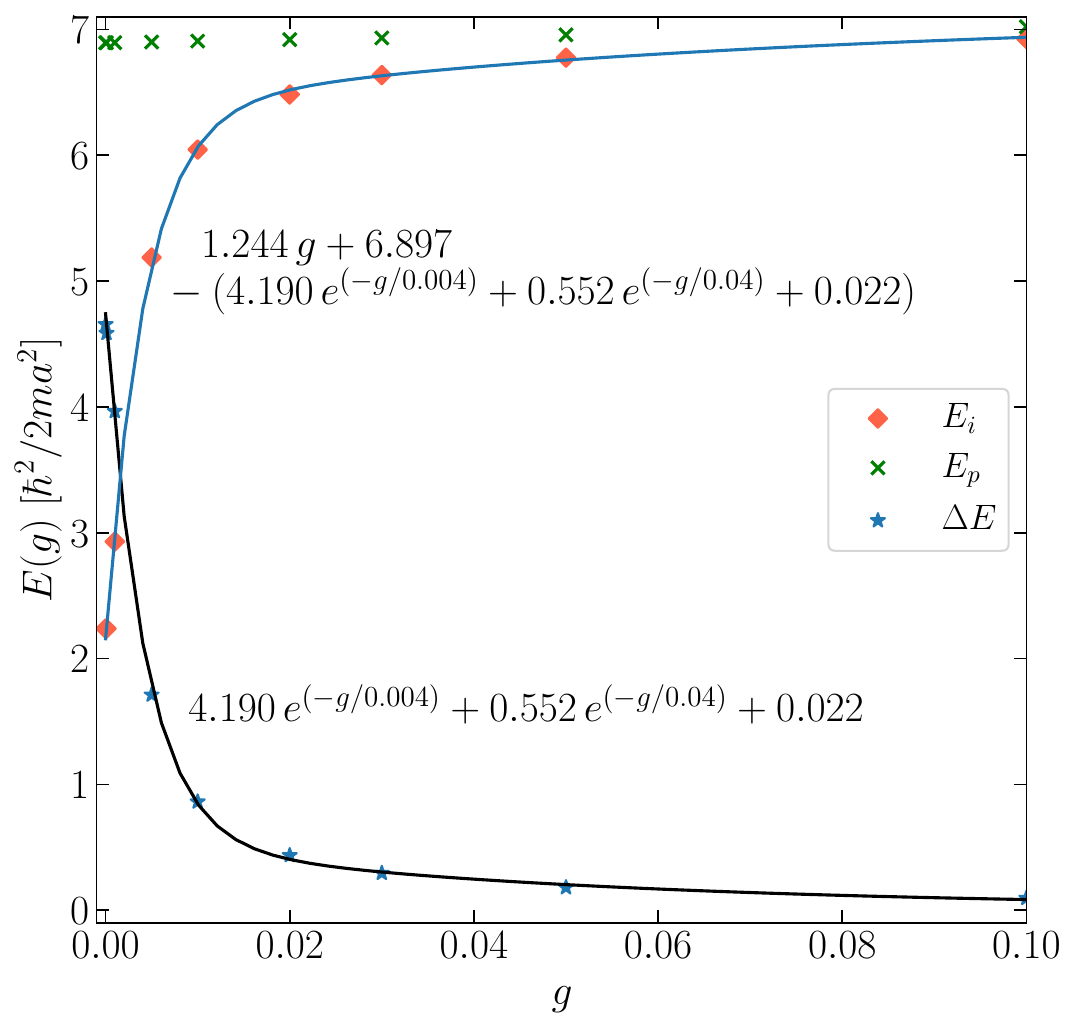}
  \caption{Enlargement of Fig. \ref{fig:Lieb} for $g \leq 0.1$, with two adjustments for the imperfect system ($E_i$) and for the difference ($\Delta E$) between the perfect system ($E_p$) and the imperfect system ($E_i$)}%
  \label{fig:LiebZoom}
\end{figure}

\subsection{Removing only one delta}
A regression of the GSE as a function of the inverse of the total deltas was made, one delta of the system is removed with $g=1$ this in order to obtain the trend when the system grows to infinity In Fig. \ref{fig:regresion} we observe the regression obtained, each of the points represents the value of the energy of the ground state for each amount of total deltas the regression shows that the GSE tends to grow as the number of total deltas increases. 
In Tables \ref{table:tabla4} and \ref{table:tabla5} we have a compilation of GSE values for the cases analyzed where everything mentioned above can be seen more clearly. With this in Fig. \ref{fig:Lieb} we show a comparison of what was obtained with the Lieb-Liniger case which serves to corroborate our data, since the free case would have to follow the trend of the LL gas, which happens.Let us note that in all cases except the free and perfect cases, between the GSE of the case with interaction 0 and the first case with the energies on we have an abrupt jump in its value. 
In addition to this we show the difference in the energy of the perfect system and the energy of the imperfect system and two fitted curves in Fig. \ref{fig:LiebZoom}, these show the trend of the energy of the imperfect system ($E_i$) and the difference ($\Delta E$), for the perfect case it did not fit because the trend is a straight line, for the other two cases we notice that after a certain value of $g$ close to 0.05 this trend also becomes a constant, in the difference we expect it to be 0, this is because the stronger the interaction, the system begins to ignore the deltas.

\section{Ground state energy, Chemical potential and distribution function}
The Lieb-Liniger model \cite{LL} describes an one-dimensional Bose gas with $N$ bosons interacting through a repulsive contact potential.
The Hamiltonian is

\begin{equation}
  {\hat{H}}_{LL} = - \frac{\hbar^2}{2m} \sum_{j=1}^{N} \frac{\partial^2}{\partial x^2_{j}} + g_{1D} \sum_{i=1}^{N} \sum_{j=i+1}^{N} \delta(z_i-z_j)
  \label{eq:Hamilton_Lieb}
\end{equation}
This Hamiltonian can be accurately diagonalized using the Bethe ansatz to obtain the energy per particle at zero temperature which can be written as
\begin{equation}
  E(n_1)/N = \frac{\hbar^2}{2m} n_1^2 e(\gamma(n_1))
  \label{eq:Energia_Lieb}
\end{equation}
where
%
%\begin{equation}
$n_1=N/L$ the number density, $ \gamma = {m \, g_{1D}}/({\hbar^2 \, n_1})$
% = \frac{-2}{a_{1D}n_1}
% \label{eq:gamma}
%\end{equation}
%
and
\begin{equation}
  e(\gamma) = \frac{\gamma^3}{\lambda^3(\gamma)} \int^{+1}_{-1} g(x, \gamma) x^2dx
  \label{eq:GammaInt}
\end{equation}
where the functions  $g(x, \gamma) $ and $\lambda(\gamma)$  satisfy the relations
\begin{equation}
  g(x, \gamma) - \frac{1}{2\pi} = \frac{\lambda(\gamma)}{\pi} \int^{+1}_{-1} \frac{g(x', \gamma)}{\lambda ^2 (\gamma)+ (x-x')^2}dx'
\end{equation}
and
\begin{equation}
  \lambda (\gamma) = \gamma \int^{+1}_{-1} g (x, \gamma) dx
\end{equation}
respectively.

With the information of the energy per particle it is easy to calculate the chemical potential with
the following derivative $\mu = \partial(n_1 E(n_1))/\partial n_1$.
% An important part that gives
%consistency to this result is that for high densities, that is, $\gamma <<1$. % we recover the limit
%$n_{1D} {a^2_{\perp}}/{a_s}>>1 $ where the mean field theory holds.
%

\section{Conclusions}
We have obtained the ground state energy of a weakly interacting Bose gas at zero temperature within perfect and defective 1D artificial crystals by numerically solving the Gross-Pitaevskii equation
using the Gradient Flow with Discrete Normalization method.  

We model the periodic structure with a Dirac Comb potential to generate a chain of scattering points like atoms in a crystal.  
Punctual vacancies in the crystal are generated by randomly removing a fraction of the total number of deltas of the Dirac Comb potential, leaving us with an imperfect crystal. 
%The size of the system was increased by increasing the original number of  deltas but keeping both the separation between contiguous deltas and the particle density such as a curve was fitted by the method of non-linear least squares to observe the behavior of the ground state energy and the chemical potential in the thermodynamic limit. 

In addition to the GSE we give the probability density function of the studied systems as well as their chemical potential.  
 
The energy per particle increases proportionally to the intensity of the deltas, with respect to the energy of the free ideal gas.

For the ideal Bose gas ($g=0$) within a perfect crystal
the GSE per particle increases proportionally to the delta strength, with respect to that of the free IBG (zero energy in the T.L.). However, when vacancies were introduced, the GSE decreases exponentially from its value per particle in the perfect crystal until it reaches that value of the free IBG when the vacancy ratio is 100 \%, i.e., all deltas have been removed. In particular, when a single delta is removed (punctual defect) breaking the translational symmetry, 
%the energy decreased from its value in the perfect case.
%relative to the perfect structure case. 
%When we have 100\% vacancies we recover the energy of the
%free cases with or without interaction.
% For a ideal Bose gas trapped by a Dirac comb potential, the absence   
%% is well studied that the presence of
%a delta (punctual defect) generates
an energy gap between the GSE and the first excited state of the particle energy spectrum is observed which promotes the Bose Einstein condensation at finite temperature for the IBG within the imperfect 1D crystal. For delta strength such that $P_0 = 10$ the gap magnitude is about 4.656 $\hbar^2/2ma^2$ and independent of initial number of deltas. 
%
%The energy magnitude of the first excited state is equal to the GSE of ideal Bose gas trapped by the perfect Dirac comb potential.

For a free interacting Bose gas, its GSE is given analytically in the beautiful work of Lieb and Liniger \cite{LL} as a function of the Lieb interaction parameter $\gamma$, which is equal to $g$ in this work;
the GSE increases almost linearly with $g$. For this same system our calculations of the GSE  fall very close to the Lieb-Liniger curve, id est, with an error less than 17\%.

When the weakly interacting Bose gas is trapped by a perfect Dirac comb potential its GSE for each value of $g$, increases proportionally to $P_0$ values. For $P_0 = 0$ we recover the GSE of the free weak interacting Bose gas while if $P_0 = 10$ the GSE increseas about 7.144 $\pm$ 0.250 $\hbar^2/2m a^2$ with respect to the case with $P_0 = 0$, within the interval of $g$ analyzed.

If the artificial crystal presents only one punctual defect the GSE of the weak interacting Bose gas presents a decrease with respect to the GSE value of the gas trapped by the perfect crystal. The energy difference is maximum for $g=0$, i.e. for the trapped IBG, which  decreases exponentially as $g$ increases to the point where $g \geq 0.2$ the energy difference of both systems shows almost the same small value (with respect to the case $g=0$) but the energy values for the case with one vacancy being always slightly lower. This is expected since as the interaction between bosons increases, the effect of a punctual vacancy on the GSE becomes imperceptible. For this case, for each value of $g$, we predict the value of the GSE for the infinite system. To do this, the GSE was plotted as a function of the inverse of the number of deltas (inverse size of the system) which we extrapolated to zero to obtain the ground state energy of the infinite system. The GSE of the infinite system is a maximum value to which the GSE of finite systems tends as we increase their size.

%The size efect on the GSE
%.greater in the interval of g values ​​between [0, 0.1] where it is observed that the difference increases exponentially until reaching the value of 4.6 e0 at g=0. 

%For values of the interaction larger than g=.2 with a punctual defect is slightly less than that of the gas trapped by the perfect potential. These energy differences are in the interval [0.001, 0.1] 
%is about $0.05 \pm 0.5   
%$\hbar^2/2ma^2$  and they are interaction magnitude dependent. 
 
%In our system of weakly interacting bosons, the imperfection also generates a jump, however this is
%not enough to determine if there is some kind of phase transition, although it could be an
%indication that it is.

For any interaction value $g$, increasing the vacancy percentage leads to an exponential decrease of the GSE, from its perfect trapped value till the GSE value for the free gas case.  At the same time, for a fixed vacancy ratio, the GSE increases almost linearly as a function of the interaction magnitude $g$ between particles.

%ABOUT THE CHEMICAL POTENTIAL.
Additionally, we observe that the chemical potential increases proportionally to the magnitude
of the interactions and that the presence of the perfect structure also increases its value. While
the imperfections cause the chemical potential value to decrease with respect to the perfect system. Also, we observe that the chemical potential values are larger than the GSE and the difference increases as a function of the interaction strength between bosons as well as with the delta strength. %With the
%extrapolation it was found that the systems with a thousand deltas only differ in significant digits
%in the value of the ground state energy, so a next step would be to analyze these systems since they
%are already very similar to infinite systems. 

%In the case of the chemical potential, what we find is that the interactions , making its value greater than that of the ground state energy.

When the artificial crystal is perfect the probability density is completely periodic, however in presence of the vacancies it shows a 
localization at the vacancy positions which is lost as the magnitude of the interaction between bosons increases. 

\section{Declarations}
Funding:   Authors thank partial financial support received from DGAPA grant PAPIIT IN114523.

Availability of data and material: None.

Competing interests: The authors declare that there is no conflict of interest in relation to the publication of this article, nor any financial interests.

Authors Contributions: Both authors contributed equally to: the conception and design of the study, numerical calculations, data analysis, and interpretation of the results. Both authors read and approved the final manuscript.

In Memoriam: This article is dedicated to the memory of our dear colleague, Dr. Omar Abel Rodríguez López (12/26/1985-03/09/2025), who tragically died in a traffic accident while we were completing this manuscript. His expertise in the numerical solution of the Gross-Pitaevskii equation was invaluable for solving the corresponding GP equation for the boson gas within the one-dimensional perfect and imperfect structures reported here.

\end{document}